\documentclass{ws-procs961x669}            

\usepackage{xspace} 
\usepackage{upgreek}

\def\lhcb {\mbox{LHCb}\xspace}

\def\lhc    {\mbox{LHC}\xspace}

\def\MagUp {\mbox{\em Mag\kern -0.05em Up}\xspace}

 \def\Ppi         {\ensuremath{\pi}\xspace}

 \def\Ppsi        {\ensuremath{\psi}\xspace}                 
                  
 \mathchardef\PDelta="7101
 \mathchardef\PXi="7104
 \mathchardef\PLambda="7103
 \mathchardef\PSigma="7106
 \mathchardef\POmega="710A
 \mathchardef\PUpsilon="7107
                  
 \def\PB      {\ensuremath{B}\xspace}                 
                  
 \def\PD      {\ensuremath{D}\xspace}

 \def\PJ      {\ensuremath{J}\xspace}                 
 \def\PK      {\ensuremath{K}\xspace}

 \def\Pb      {\ensuremath{b}\xspace}                 
 \def\Pc      {\ensuremath{c}\xspace}

 \def\Pi      {\ensuremath{i}\xspace}

 \def\Pp      {\ensuremath{p}\xspace}                 
 \def\Pq      {\ensuremath{q}\xspace}

\DeclareRobustCommand{\optbar}[1]{\shortstack{{\miniscule (\rule[.5ex]{1.25em}{.18mm})}
  \\ [-.7ex] $#1$}}

\def\quark     {{\ensuremath{\Pq}}\xspace}
\def\quarkbar  {{\ensuremath{\overline \quark}}\xspace}
\def\qqbar     {{\ensuremath{\quark\quarkbar}}\xspace}

\def\cquark    {{\ensuremath{\Pc}}\xspace}

\def\bquark    {{\ensuremath{\Pb}}\xspace}

\def\pion   {{\ensuremath{\Ppi}}\xspace}

\def\pip    {{\ensuremath{\pion^+}}\xspace}
\def\pim    {{\ensuremath{\pion^-}}\xspace}

\def\kaon    {{\ensuremath{\PK}}\xspace}
  \def\Kbar    {{\kern 0.2em\overline{\kern -0.2em \PK}{}}\xspace}

\def\KorKbar    {\kern 0.18em\optbar{\kern -0.18em K}{}\xspace}

\def\Kp      {{\ensuremath{\kaon^+}}\xspace}
\def\Km      {{\ensuremath{\kaon^-}}\xspace}

\def\Kstarz  {{\ensuremath{\kaon^{*0}}}\xspace}

  \def\Dbar    {{\kern 0.2em\overline{\kern -0.2em \PD}{}}\xspace}

\def\DorDbar    {\kern 0.18em\optbar{\kern -0.18em D}{}\xspace}

\def\Bbar    {{\ensuremath{\kern 0.18em\overline{\kern -0.18em \PB}{}}}\xspace}

\def\BorBbar    {\kern 0.18em\optbar{\kern -0.18em B}{}\xspace}

\def\jpsi     {{\ensuremath{{\PJ\mskip -3mu/\mskip -2mu\Ppsi\mskip 2mu}}}\xspace}

  \def\Y#1S{\ensuremath{\PUpsilon{(#1S)}}\xspace}

\def\proton      {{\ensuremath{\Pp}}\xspace}

\def\Deltares    {{\ensuremath{\PDelta}}\xspace}

\def\Xires       {{\ensuremath{\PXi}}\xspace}

\def\Lz          {{\ensuremath{\PLambda}}\xspace}
\def\Lbar        {{\ensuremath{\kern 0.1em\overline{\kern -0.1em\PLambda}}}\xspace}
\def\LorLbar    {\kern 0.18em\optbar{\kern -0.18em \PLambda}{}\xspace}

\def\Sigmares    {{\ensuremath{\PSigma}}\xspace}

\def\Lb      {{\ensuremath{\Lz^0_\bquark}}\xspace}

\def\Lc      {{\ensuremath{\Lz^+_\cquark}}\xspace}

\def\Xim    {{\ensuremath{\Xires^-}}\xspace}
\def\Xicz    {{\ensuremath{\Xires^0_\cquark}}\xspace}
\def\Xicp    {{\ensuremath{\Xires^+_\cquark}}\xspace}

\def\Sigmap    {{\ensuremath{\Sigmares^+}}\xspace}

\def\BF         {{\ensuremath{\cal B}}\xspace}

\def\BR         {\BF}
\newcommand{\decay}[2]{\ensuremath{#1\!\to #2}\xspace}         

\def\to                 {\ensuremath{\rightarrow}\xspace}

\def\CP                {{\ensuremath{C\!P}}\xspace}
\def\CPT               {{\ensuremath{C\!PT}}\xspace}

\newcommand{\ATbar}{{\ensuremath{\kern 0.1em\overline{\kern -0.1em A}_{\widehat{T}}}}\xspace}

\newcommand{\CTbar}{{\ensuremath{\kern 0.1em\overline{\kern -0.1em C}_{\widehat{T}}}}\xspace}

\def\C#1      {\ensuremath{\mathcal{C}_{#1}}\xspace}                       
\def\Cp#1     {\ensuremath{\mathcal{C}_{#1}^{'}}\xspace}                    
\def\Ceff#1   {\ensuremath{\mathcal{C}_{#1}^{\mathrm{(eff)}}}\xspace}        
\def\Cpeff#1  {\ensuremath{\mathcal{C}_{#1}^{'\mathrm{(eff)}}}\xspace}       
\def\Ope#1    {\ensuremath{\mathcal{O}_{#1}}\xspace}                       
\def\Opep#1   {\ensuremath{\mathcal{O}_{#1}^{'}}\xspace}                    

\newcommand{\tev}{\ensuremath{\mathrm{\,Te\kern -0.1em V}}\xspace}
\newcommand{\gev}{\ensuremath{\mathrm{\,Ge\kern -0.1em V}}\xspace}
\newcommand{\mev}{\ensuremath{\mathrm{\,Me\kern -0.1em V}}\xspace}
\newcommand{\kev}{\ensuremath{\mathrm{\,ke\kern -0.1em V}}\xspace}
\newcommand{\ev}{\ensuremath{\mathrm{\,e\kern -0.1em V}}\xspace}
\newcommand{\gevc}{\ensuremath{{\mathrm{\,Ge\kern -0.1em V\!/}c}}\xspace}
\newcommand{\mevc}{\ensuremath{{\mathrm{\,Me\kern -0.1em V\!/}c}}\xspace}
\newcommand{\gevcc}{\ensuremath{{\mathrm{\,Ge\kern -0.1em V\!/}c^2}}\xspace}
\newcommand{\gevgevcccc}{\ensuremath{{\mathrm{\,Ge\kern -0.1em V^2\!/}c^4}}\xspace}
\newcommand{\mevcc}{\ensuremath{{\mathrm{\,Me\kern -0.1em V\!/}c^2}}\xspace}

\def\m    {\ensuremath{\rm \,m}\xspace}

\def\cm   {\ensuremath{\rm \,cm}\xspace}

\def\mub{\ensuremath{{\rm \,\upmu b}}\xspace}

\def\invfb   {\ensuremath{\mbox{\,fb}^{-1}}\xspace}

\def\gsim{{~\raise.15em\hbox{$>$}\kern-.85em
          \lower.35em\hbox{$\sim$}~}\xspace}
\def\lsim{{~\raise.15em\hbox{$<$}\kern-.85em
          \lower.35em\hbox{$\sim$}~}\xspace}

\def\tell1  {TELL1\xspace}
\def\ukl1   {UKL1\xspace}

\newcommand{\ie}{\mbox{\itshape i.e.}\xspace}

\newcommand{\effCH}{{\ensuremath{\varepsilon_{\rm CH}}}\xspace}
\newcommand{\effDF}{{\ensuremath{\varepsilon_{\rm DF}}}\xspace}
\newcommand{\effdet}{{\ensuremath{\varepsilon_{\rm det}}}\xspace}
\newcommand{\effgeo}{{\ensuremath{\varepsilon_{\rm geo}}}\xspace}
\newcommand{\efftrigger}{{\ensuremath{\varepsilon_{\rm trigger}}}\xspace}
\newcommand{\efftrack}{{\ensuremath{\varepsilon_{\rm track}}}\xspace}

\begin{document}
\title{Search for new physics via baryon EDM at LHC}

\author{L. Henry$^1$, D. Marangotto$^2$, A. Merli$^{2,3}$, N. Neri$^{2,3}$, J. Ruiz$^1$, F. Martinez Vidal$^1$}

\address{$^1$IFIC, Universitat de Val\`encia-CSIC, Valencia, Spain\\
$^2$INFN Sezione di Milano and Universit\`a di Milano, Milan, Italy\\
$^3$CERN, Geneva, Switzerland}

\begin{abstract}
Permanent electric dipole moments (EDMs) of fundamental particles provide powerful probes
for physics beyond the Standard Model. We propose to search for the EDM of strange
and charm baryons at LHC, extending the ongoing experimental program on the neutron, muon, atoms,
molecules and light nuclei.
The EDM of strange \Lz baryons, selected from weak decays of charm
baryons produced in \proton\proton collisions at LHC,
can be determined by studying the spin precession in the magnetic field of the detector tracking system.
A test of \CPT symmetry can be performed by measuring the magnetic dipole moment of \Lz and \Lbar baryons.
For short-lived \Lc and \Xicp baryons, to be produced in a fixed-target experiment using the 7 \tev LHC beam
and channeled in a bent crystal, the spin precession is induced by the intense
electromagnetic field between crystal atomic planes.
The experimental layout based on the \lhcb detector and the expected sensitivities in the coming years are discussed.
\end{abstract}


\bodymatter

\section{Introduction}
\label{sec:intro}

The magnetic dipole moment (MDM) and the electric dipole moment (EDM) are static 
properties of particles that determine the spin motion in an external electromagnetic field, as
 described by the T-BMT equation~\cite{Thomas:1926dy,Thomas:1927yu,Bargmann:1959gz}. 

The EDM is the only static property of a particle
that requires the violation of parity ($P$) and time reversal ($T$) symmetries
and thus, relying on \CPT invariance, the violation of \CP symmetry.
The amount of \CP violation in the weak interactions of quarks is not sufficient to explain
the observed 
imbalance between matter and antimatter in the Universe.
CP-violation in strong interactions is strongly bounded by the experimental limit on the neutron EDM~\cite{Afach:2015sja}.
In the Standard Model (SM), contributions to the EDM of baryons are highly
suppressed but can be largely enhanced in some of its extensions.
Hence, the experimental searches for the EDM of fundamental
particles provide powerful probes for 
physics beyond the SM.

Since EDM searches started in the fifties~\cite{Purcell:1950zz,Smith:1957ht},
there has been an intense experimental program,
leading to limits on the EDM of leptons~\cite{Baron:2013eja,Bennett:2008dy,Inami:2002ah},
neutron~\cite{Afach:2015sja}, heavy atoms~\cite{Griffith:2009zz},
proton (indirect from $^{199}$Hg)~\cite{Dmitriev:2003sc}, and \Lz baryon~\cite{Pondrom:1981gu}.
New experiments are ongoing and others are planned, including those based on storage rings
for muon~\cite{Grange:2015fou,Saito:2012zz}, proton and light nuclei~\cite{Anastassopoulos:2015ura,Pretz2015JEDI,Khriplovich:1998zq}. Recently we proposed to improve the limit on strange baryons and extend it to charm and bottom baryons~\cite{Botella:2016ksl,Bagli:2017foe}. 

EDM searches of fundamental particles rely on the measurement of the spin precession angle
induced by the interaction with the electromagnetic field.
For unstable particles this is challenging since the 
precession has to take place before the decay. A solution to this problem requires large samples of
high energy 
polarized particles traversing an intense electromagnetic field.

Here we reviewed the unique possibility to search for the EDM of the strange \Lz baryon
and of the charmed baryons at LHC. 
Using the experimental upper limit of the neutron EDM, the absolute value of the \Lz EDM is predicted
to be \mbox{$<4.4\times 10^{-26}~e\cm$~\cite{Guo:2012vf,Atwood:1992fb,Pich:1991fq,Borasoy:2000pq}}, while 
the indirect constraints
on the charm EDM are weaker, 
$\lsim 4.4\times 10^{-17}~e\cm$~\cite{Sala:2013osa}. 
Any experimental observation of an EDM 
would indicate a new source of \CP violation 
from physics beyond the SM.
The EDM of the long-lived \Lz baryon was measured to be
$<1.5 \times 10^{-16}~e\cm$ (95\% C.L.) in a fixed-target experiment
at Fermilab~\cite{Pondrom:1981gu}.
No experimental measurements exist for short-lived charm baryons
since negligibly small spin precession would be induced by
magnetic fields used in current particle detectors.

\section{Experimental setup}

The magnetic and electric dipole moment of a spin-1/2 particle is given (in Gaussian units) by
$\bm{\mu} = g \mu_B {\mathbf s}/2$ and \mbox{$\bm{ \delta} = d \mu_B {\mathbf s}/2$}, respectively,
where $\mathbf{s}$ is the spin-polarization vector\footnote{The spin-polarization vector
is defined such as 
$\mathbf s = 2 \langle \mathbf S \rangle / \hbar$, 
where $\mathbf S$ is the spin operator.
}
and $\mu_B=e \hbar / (2 m c)$ is the particle magneton, with $m$ its mass.
The $g$ and $d$ dimensionless factors 
are also referred to as the gyromagnetic and gyroelectric ratios.
The experimental setup to measure the change of the spin direction in an elextromagnetic field  relies 
on three main elements:
\begin{romanlist}
\item a source of polarized particles whose direction 
and polarization degree are known;
\item an intense electromagnetic field able to induce a sizable spin precession angle during 
the lifetime of the particle;
\item the detector to measure the final polarization vector
by analysing the angular distribution of the particle decays.
\end{romanlist}

\subsection{\Lz and \Lbar case}

Weak decays of heavy baryons (charm and beauty), mostly produced in the
forward/backward directions at \lhc, 
can induce large longitudinal polarization due to parity violation. 
For example, the decay of unpolarized \Lc baryons to the $\Lz\pip$ final state~\cite{Link:2005ft}, 
produces \Lz baryons with longitudinal polarization $\approx -90\%$.
Another example is the $\Lb\to\Lz\jpsi$ decay where \Lz baryons are produced
almost 100\% longitudinally polarized~\cite{Aaij:2013oxa,Aad:2014iba}.

The spin-polarization vector $\mathbf s$ of an ensemble of \Lz baryons
can be analysed through the angular distribution of the $\Lz\to \proton \pim$ decay~\cite{Lee:1957qs,Richman1984},
\begin{equation}
\label{eq:AngDist}
\frac{dN}{d\Omega'} \propto 1 + \alpha \mathbf s \cdot \hat{\mathbf k} ~,
\end{equation}
where $\alpha = 0.642 \pm 0.013$~\cite{Olive:2016xmw} is the decay asymmetry parameter. 
The \CP invariance in the \Lz decay implies $\alpha = -\overline{\alpha}$, where $\overline\alpha$ is the decay parameter of 
the charge-conjugate decay.
The unit vector $\hat {\mathbf k} = (\sin\theta'\cos\phi', \sin\theta'\sin\phi', \cos\theta')$
indicates the momentum direction  of the proton in the \Lz 
helicity frame,
with $\Omega' = (\theta',\phi')$ the corresponding solid angle.
For the particular case of \Lz
flying along the $z$ axis in the laboratory frame,
an initial longitudinal polarization $s_0$, \ie $\mathbf s_0=(0,0,s_0)$,
and $\mathbf B = (0,B_y,0)$, 
the solution of the T-BMT equation is~\cite{Botella:2016ksl}
\begin{equation}
\label{eq:sSimpleCase}
\mathbf s =
\begin{cases}
s_x = - s_{0} \sin\Phi  \\
s_y = - s_{0} \dfrac{d \beta }{g} \sin\Phi \\
s_z =   s_{0} \cos\Phi \\
\end{cases}
\end{equation}
where $\Phi = \frac{D_y\mu_B}{\beta \hbar c} \sqrt{d^2 \beta^2 + g^2}\approx\frac{g D_y \mu_B}{\beta \hbar c}$
with $D_y\equiv D_y(l) = \int_0^l B_y dl'$ the integrated magnetic field
along the \Lz flight path. 
The polarization vector precesses in the $xz$ plane, normal to the magnetic field, with 
the precession angle $\Phi$ proportional to the 
gyromagnetic factor of the particle.
The presence of an EDM introduces a non-zero
$s_y$ component perpendicular to the precession plane of 
the MDM, otherwise not present.
At LHCb, with a tracking dipole magnet providing an integrated field $D_y \approx \pm 4~\mathrm{T \m}$~\cite{LHCb-DP-2014-002}, 
the maximum precession angle for particles traversing the entire magnetic field region yields $\Phi_{\rm max} \approx \pm \pi/4$, 
and allows to achieve about 70\% of the maximum $s_y$ component.
Moreover, 
a test of \CPT symmetry can be performed by comparing the $g$ and $-\bar g$ factors for \Lz and
\Lbar baryons, respectively, which precess in opposite directions as 
$g$ and $d$ change sign from particle to antiparticle.

\subsection{Charm baryon case}

The \Lc baryon EDM can be extracted by measuring the precession of the
polarization vector of channeled particles in a bent crystal.
There, a positively-charged particle channeled between atomic planes moves along a curved path under
the action of the intense electric field between crystal planes. In the instantaneous rest frame of the particle
the electromagnetic field causes the spin rotation.
The signature of the EDM
is a polarization component perpendicular to the initial baryon
momentum and polarization vector, otherwise not present, similarly to the case of
the \Lz baryon.

The phenomenon of spin precession of positively-charged particles channeled in a bent crystal
was firstly observed by the E761 collaboration that measured the MDM
of the strange \Sigmap baryon~\cite{Chen:1992wx}.
The feasibility of the measurement at \lhc\ energies
offers clear advantages with respect to lower beam energies since the estimated number
of channeled charm baryons is proportional to $\gamma^{3/2}$, where $\gamma$ is the Lorentz factor of the particles~\cite{Baryshevsky:2016cul}. 

\begin{figure*}[htb]
\centering
{ \includegraphics[width=0.34\linewidth]{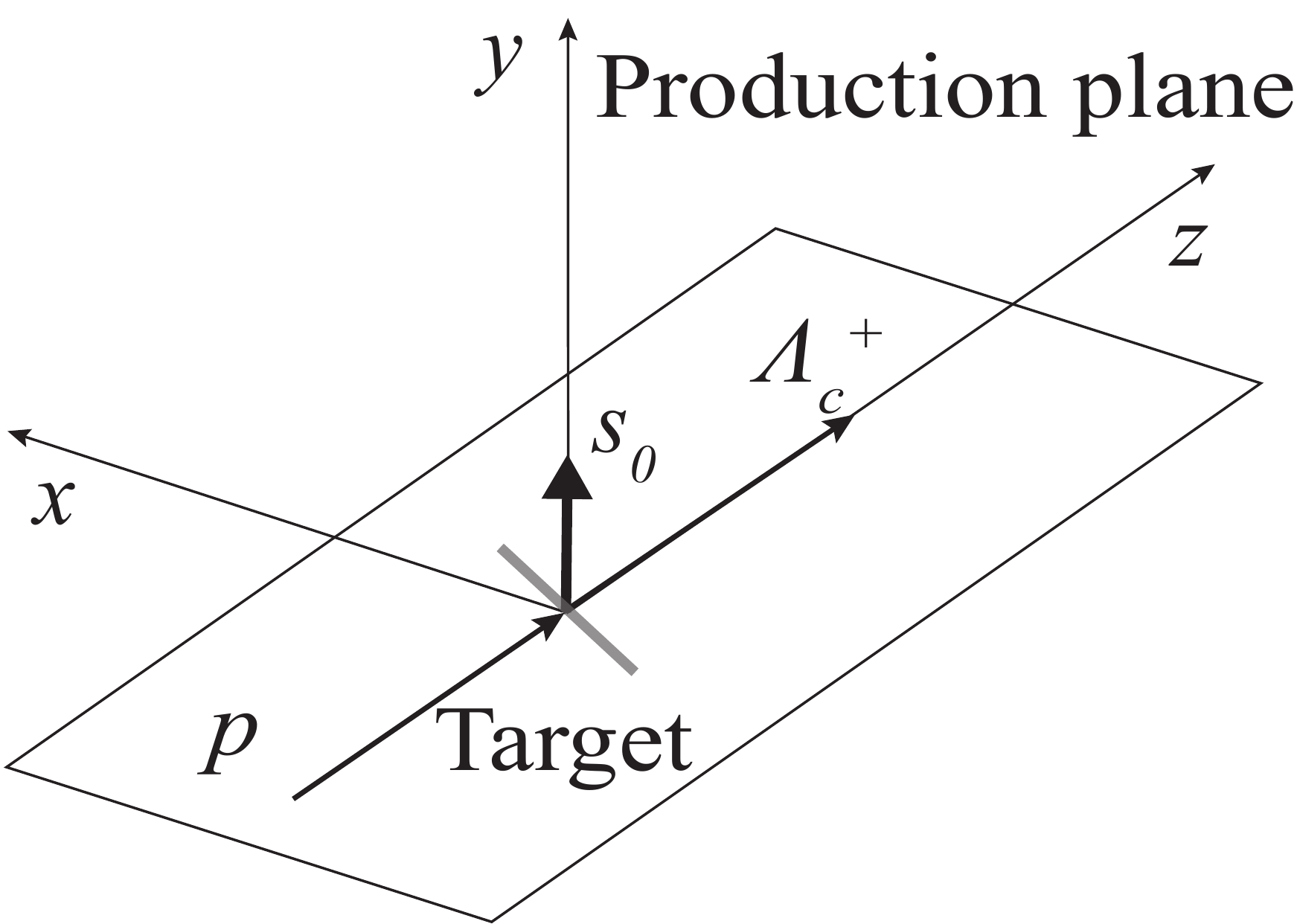} }
\hspace{1cm}    
{ \includegraphics[width=0.54\linewidth]{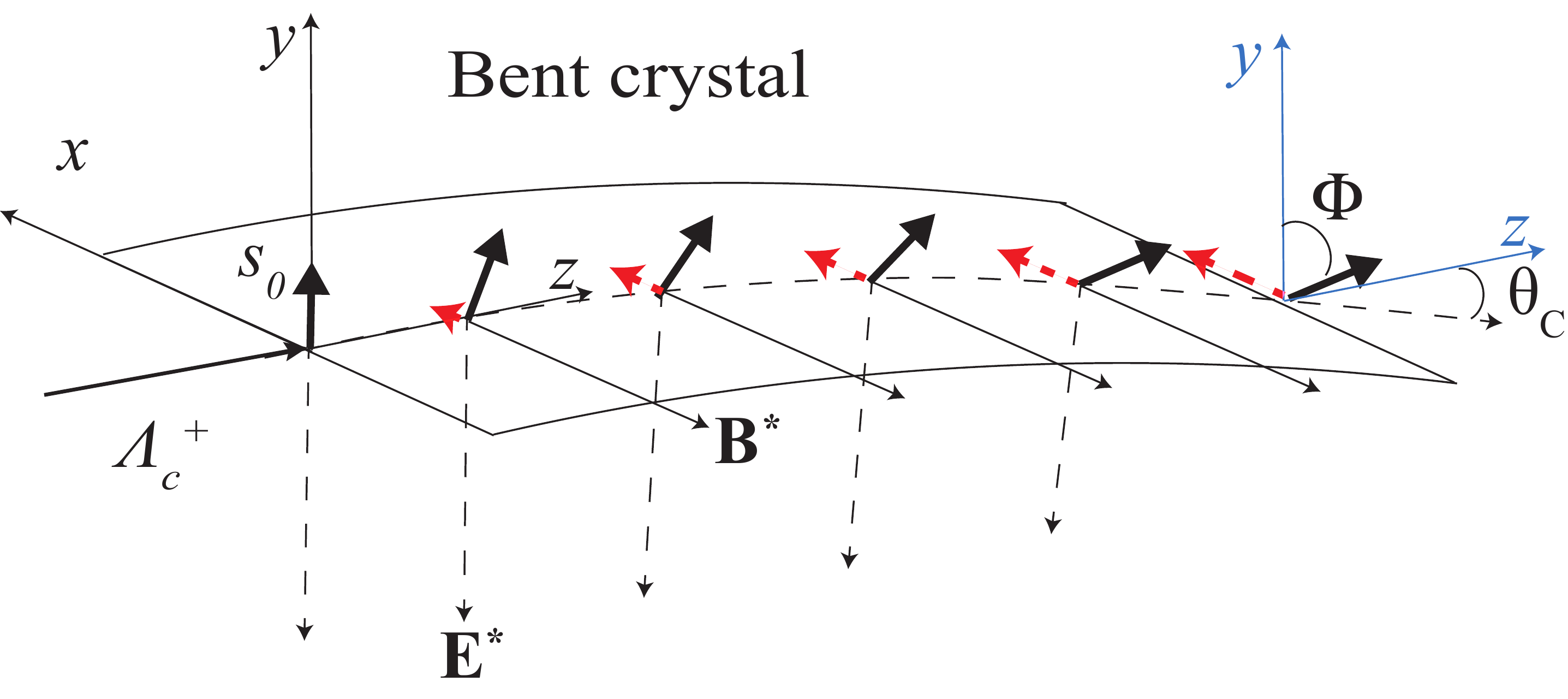} }
\caption{(Left) Production plane of the \Lc baryon defined by the proton and the \Lc momenta.
  The initial polarization vector $\mathbf s_0$ is perpendicular to the production plane,
  along the $y$ axis, due to parity conservation in strong interactions~\cite{Jacob:1959at}.
  (Right) Deflection of the baryon trajectory and
  spin precession in the $yz$ and $xy$ plane induced by the MDM and the EDM, respectively.
  The red (dashed) arrows indicate the (magnified) $s_x$ spin component
  proportional to the particle EDM. $\Phi$ is the MDM precession angle and $\theta_C$ is
 the crystal bending angle. $\mathbf{E^*}$ and $\mathbf{B^*}$ are the intense electromagnetic field in the particle rest frame~\cite{Baryshevsky:2015zba,Kim:1982ry} which induce spin precession.}
\label{fig:Lc_ProdBending}
\end{figure*}
In the limit of large boost with Lorentz factor $\gamma \gg 1$, the precession angle $\Phi$, shown in \figurename~\ref{fig:Lc_ProdBending},
induced by the MDM is~\cite{Lyuboshits:1979qw}
\begin{equation}
\Phi \approx \frac{g-2}{2}\gamma \theta_C,
\label{eq:MDM_angle}
\end{equation}
where $g$ is the gyromagnetic factor, $\theta_C=L/\rho_0$ is the crystal bending angle, $L$ is the circular arc of the crystal and $\rho_0$ the curvature radius .

In presence of a non-zero EDM, the spin precession is no longer confined to the $yz$ plane,
originating a $s_x$ component proportional
to the particle EDM represented by the red (dashed) arrows in (Right) \figurename~\ref{fig:Lc_ProdBending}.
The polarization vector, after channeling through the crystal is~\cite{Botella:2016ksl}
\begin{equation}
\mathbf s ~=~
\left\lbrace
\begin{array}{l}
s_{x} \approx   s_0 \dfrac{d}{g-2}  (\cos{\Phi}-1)  \\
s_{y} \approx   s_{0} \cos\Phi \\
s_{z} \approx   s_{0} \sin\Phi
\end{array}
\right. ,
\label{eq:EDM_LcPol}
\end{equation}
where $\Phi$ is given by Eq.~(\ref{eq:MDM_angle}).
The MDM and EDM information can be extracted from the measurement of the spin polarization of channeled baryons at the exit of the crystal, via the study of the angular distribution of final state particles. For \Lc decaying to two-body final states such as $f = \Delta^{++}\Km, \proton\Kstarz, \Delta(1520)\pip$ and $\Lambda\pim$, the angular distribution is described by Eq.~\ref{eq:AngDist}. A Dalitz plot analysis would provide the ultimate sensitivity to the EDM measurement.

The initial polarization $s_0$ would require in principle the measurement of the angular distribution for unchanneled baryons. In practice this is not required since the measurement of the three components of the final polarization vector for channeled baryons allows a simultaneous determination of $g, d$ and $s_0$, up to discrete ambiguities. These can be solved exploiting the dependence of the angular distribution with the \Lc boost $\gamma$, as discussed in Ref.~\cite{Bagli:2017foe}.

\section{Sensitivity studies}

\subsection{\Lz and \Lbar case}

The number of \Lz particles produced can be estimated as 
\begin{equation}
N_\Lz= 2 \mathcal{L} \sigma_\qqbar f(\quark \to H)\BR(H \to \Lz X') \BR(\Lz\to\proton\pim)\BR(X'\to\mathrm{charged}) , 
\end{equation}
where $\mathcal{L}$ is the total integrated luminosity, $\sigma_\qqbar$ ($\quark=\cquark,\bquark$) are the heavy quark 
production cross sections from \proton\proton collisions at $\sqrt{s}=13$\tev~\cite{Aaij:2015bpa,FONLLWEB,Aaij:2010gn,Aaij:2015rla},
and $f$ is the fragmentation fraction into the heavy 
baryon $H$~\cite{Lisovyi:2015uqa,Gladilin:2014tba,Amhis:2016xyh,Galanti:2015pqa}.
\begin{table}
\tbl{Dominant \Lz production mechanisms from heavy baryon decays and estimated yields
  produced per \invfb at $\sqrt{s}=13$\tev,
shown separately for SL and LL topologies.
The \Lz baryons from \Xim decays, produced promptly in 
the \proton\proton collisions, are given in terms of the unmeasured production cross section. 
}
{\begin{tabular}{@{}lclc@{}}
\toprule
SL events &  $N_{\Lz}/\invfb~(\times 10^{10})$  & LL events, $\Xim\to\Lz\pim$ &  $N_{\Lz}/\invfb~(\times 10^{10})$  \\ 
\colrule
$\Xicz\to\Lz\Km\pip$       & 7.7 & $\Xicz\to\Xim\pip\pip\pim$ & 23.6 \\
$\Lc\to\Lz\pip\pip\pim$    & 3.3 & $\Xicz\to\Xim\pip$         & 7.1 \\
$\Xicp\to\Lz\Km\pip\pip$   & 2.0 & $\Xicp\to\Xim\pip\pip$     & 6.1 \\
$\Lc\to\Lz\pip$            & 1.3 & $\Lc\to\Xim\Kp\pip$        & 0.6 \\
$\Xicz\to\Lz\Kp\Km$ (no $\phi$)  & 0.2 & $\Xicz\to\Xim\Kp$              & 0.2 \\
$\Xicz\to\Lz\phi(\Kp\Km)$  & 0.1 & Prompt $\Xim$              & $0.13\times\sigma_{\proton\proton\to\Xim}~[\mu \rm b]$ \\
\botrule
\end{tabular}}
\label{tab:LambdaChannels}
\end{table}
In \tablename~\ref{tab:LambdaChannels} the dominant production channels and the estimated
 yields are summarised. Only the decays where it is experimentally possible to determine the production and decay vertex of the \Lz are considered.
Overall, there are about $1.5\times 10^{11}$ \Lz baryons per \invfb produced directly from heavy baryon decays 
(referred hereafter as short-lived, or SL events), 
and $3.8\times 10^{11}$ from charm baryons decaying through an intermediate \Xim particle (long-lived, or LL events).
The yield of \Lz baryons experimentally available can then be evaluated as
$N_\Lz^{\rm reco} = \epsilon_{\rm geo} \epsilon_{\rm trigger} \epsilon_{\rm reco} N_\Lz$,
where $\epsilon_{\rm geo}$, $\epsilon_{\rm trigger}$ and $\epsilon_{\rm reco}$ are
the geometric, trigger and reconstruction efficiencies of the detector system.
The geometric efficiency for SL topology has been estimated to be about 16\% using a 
Monte Carlo simulation of $\proton\proton$ collisions at $\sqrt{s}=13$\tev and the decay of heavy hadrons. 

\begin{figure*}[htb]
\centering
{ \includegraphics[width=0.48\linewidth]{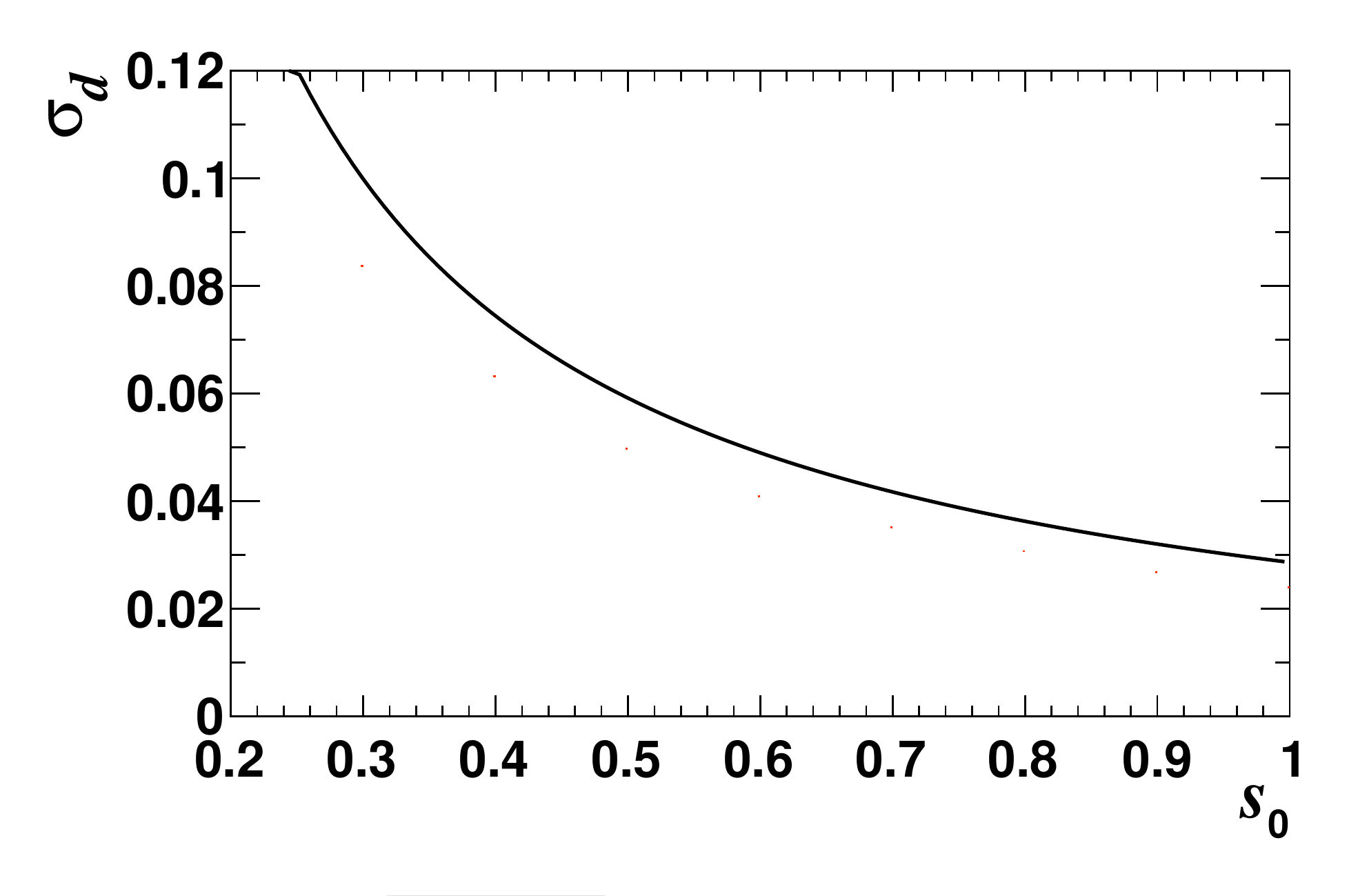} }
{ \includegraphics[width=0.48\linewidth]{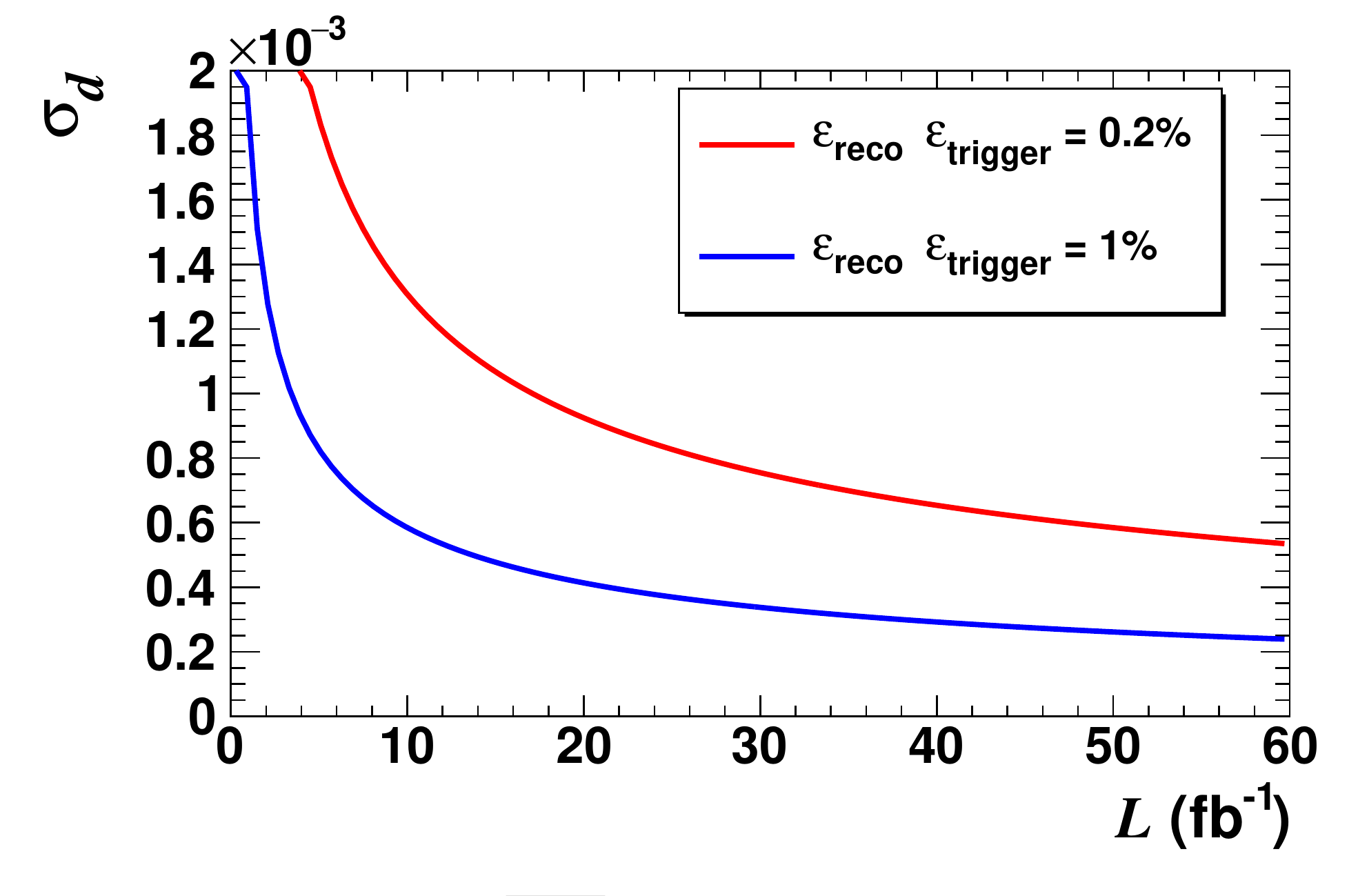} }
\caption{(Left) Dependence of the \Lz gyroelectric factor uncertainty with the initial polarization for $N_\Lz^{\rm reco}=10^6$ events, 
and (Right) as a function 
of the integrated luminosity
assuming reconstruction efficiency of  0.2\% and 1\%.}
\label{fig:Lambda_sensitivity}
\end{figure*}

To assess the EDM sensitivity, pseudo-experiments have been generated using a simplified detector geometry
that includes an approximate \lhcb magnetic field mapping~\cite{LHCb-DP-2014-002,Hicheur:2007jfk}. \Lz baryons decaying towards the end of the magnet provide most of the sensitivity to the EDM and MDM, since a sizeable spin precession could happen.
The decay angular distribution and spin dynamics have been simulated 
using Eq.~(\ref{eq:AngDist}) and the general solution as a function of the \Lz flight length~\cite{Botella:2016ksl}, respectively.
For this study the initial polarization vector $\mathbf s_0 = (0,0,s_0)$, 
with $s_0$ varying between 20\% and 100\%, and factors $g=-1.458$~\cite{Olive:2016xmw} and $d=0$,
were used. Each generated sample was fitted using an unbinned maximum likelihood method with $d$, 
$g$ and $\mathbf s_0$ 
as free parameters. The $d$-factor uncertainty 
scales with the number of events $N_\Lz^{\rm reco}$ and the initial longitudinal polarization $s_0$ as
$\sigma_d \propto 1/(s_0 \sqrt{N_\Lz^{\rm reco}} )$. 
The sensitivity saturates at large values of $s_0$, as shown in (Left) Fig.~\ref{fig:Lambda_sensitivity},
and it partially relaxes 
the requirements on the initial polarizations.
Similarly, (Right) Fig.~\ref{fig:Lambda_sensitivity} shows the expected sensitivity on the EDM as a function 
of the integrated luminosity, summing together SL and LL events, assuming global trigger
and reconstruction efficiency $\epsilon_{\rm trigger} \epsilon_{\rm reco}$ 
of 1\% (improved \lhcb software-based trigger and tracking for the upgrade detector~\cite{LHCb-TDR-016,LHCb-TDR-015}) 
and 0.2\% (current detector~\cite{LHCb-DP-2014-002}), where the efficiency estimates are based on a educated guess.
An equivalent sensitivity is obtained for the gyromagnetic factor.
Therefore, with 8~\invfb a sensitivity $\sigma_d \approx 1.5\times 10^{-3}$ could be achieved (current detector), 
to be compared to the present limit, $1.7\times 10^{-2}$~\cite{Pondrom:1981gu}. 
With 50~\invfb (upgraded detector) the sensitivity on the gyroelectric factor can reach $\approx 3\times 10^{-4}$.

\subsection{Charm baryon case}

We propose to search for charm baryon EDMs in a dedicated fixed-target experiment at the LHC to be
installed in front of the \lhcb detector. 
The target should
be attached to the crystal to maximize the yield of short-lived charm baryons to be channeled.
The rate of \Lc baryons produced with 7\tev protons on a fixed target can be estimated as
\begin{equation}
\frac{dN_\Lc}{dt} = \frac{F}{A}\sigma(\proton\proton\rightarrow\Lc X) N_T ,
\end{equation}
where $F$ is the proton rate, $A$ the beam transverse area, $N_T$ the number of target nucleons, and
$\sigma(\proton\proton\rightarrow\Lc X)$ is the cross-section for \Lc production in \proton\proton interactions at
 $\sqrt{s}=114.6\gev$ center-of-mass energy.
The number of target nucleons is $N_T=N_A\rho A T A_N/A_T $,
where $N_A$ is the Avogadro number, $\rho$ ($T$) is the target density (thickness),
and $A_T$ ($A_N$) is the atomic mass (atomic mass number). For our estimates we consider a target of tungsten thick $T=0.5\cm$ with density $\rho=19.25 {\rm \, g/cm}$.
The rate of \Lc particles channeled in the bent crystal and reconstructed in the \lhcb detector is estimated as
\begin{equation}
\label{eq:NLcReco}
  \frac{dN_\Lc^{\rm reco}}{dt} = \frac{dN_\Lc}{dt} \BR(\Lc\to f)\effCH\effDF(\Lc)\effdet ,
\end{equation}
where $\BR(\Lc\to f)$ is the branching fraction of \Lc decaying to $f$, \effCH is the efficiency of channeling \Lc inside the crystal, \effDF(\Lc) is the fraction of \Lc decaying after the crystal and \effdet is the efficiency to reconstruct the decays.
A 6.5\tev proton beam was extracted from the LHC beam halo by channeling protons in
bent crystals~\cite{Scandale:2016krl}. A beam with intensity of $5\times 10^8~\text{proton/s}$,
to be directed on a fixed target, is attainable with this technique~\cite{Lansberg:2012wj}.

The \Lc cross section is estimated from the total charm production cross section~\cite{Adare:2006hc}, rescaled to $\sqrt{s} = 114.6 \gev$ assuming a linear dependence on $\sqrt{s}$, and \Lc fragmentation function~\cite{Amhis:2016xyh} to be $\sigma_{\Lc} \approx 18.2 \mub$, compatible with theoretical predictions~\cite{Kniehl:2005de}.

The channeling efficiency in silicon crystals, including both channeling angular acceptance and dechanneling effects, is estimated to be $\effCH\approx 10^{-3}$~\cite{Biryukov1997}, while
the fraction of \Lc baryons decaying after the crystal is $\effDF(\Lc)\approx 19\%$, for 
$\gamma = 1000$ and 10\cm crystal length.
The geometrical acceptance for $\Lc\to\proton\Km\pip$ decaying into the \lhcb detector is $\effgeo \approx 25\%$ according
to simulation studies. 
 The \lhcb software-based trigger for the upgrade detector~\cite{LHCb-TDR-016} is expected to have efficiency for charm hadrons comparable
 to the current high level trigger~\cite{LHCb-DP-2014-002}, \ie $\efftrigger \approx 80\%$. 
The tracking efficiency is estimated to be $70\%$ per track, leading to an efficiency $\efftrack \approx 34\%$ for
a \Lc decay with three charged particles.
 The detector reconstruction efficiency, $\effdet = \effgeo\efftrigger\efftrack$, is estimated to be $\effdet(\proton\Km\pip) \approx 5.4\times10^{-2}$ for \decay{\Lc}{\proton\Km\pip} decays.

Few \Lc decay asymmetry parameters $\alpha_f$ are known. 
At present,
they can be computed from existing $\Lc\to\proton\Km\pip$ amplitude analysis results~\cite{Aitala:1999uq}
yielding
$\alpha_{\Deltares^{++}\Km} = -0.67 \pm 0.30$
for the \decay{\Lc}{\Deltares^{++}\Km} decay~\cite{Botella:2016ksl}. 

For the sensitivity studies we assume $s_0=0.6$ and $(g-2)/2 = 0.3$, according to experimental results and available theoretical predictions,
respectively, quoted in Ref.~\cite{Samsonov:1996ah}.
The $d$ and $g-2$ values and errors can be derived from Eq.~\eqref{eq:EDM_LcPol}.
The estimate assumes negligibly small uncertainties on $\theta_C$, $\gamma$.

Given the estimated quantities we obtain $dN^{reco}_{\Lc}/dt  \approx 5.9 \times 10^{-3}~{\rm \,s^{-1}} = 21.2~{\rm \,h^{-1}}
$ for \decay{\Lc}{\Deltares^{++}\Km}. 
A data taking of 1 month will be sufficient to reach a sensitivity of $\sigma_\delta = 1.3 \times 10^{-17}$ on the \Lc EDM.
Therefore, a
measurement of \Lc EDM is feasible 
in \Lc quasi two-body decays at \lhcb. 

\begin{figure*}
\centering
\includegraphics[width=0.49\textwidth]{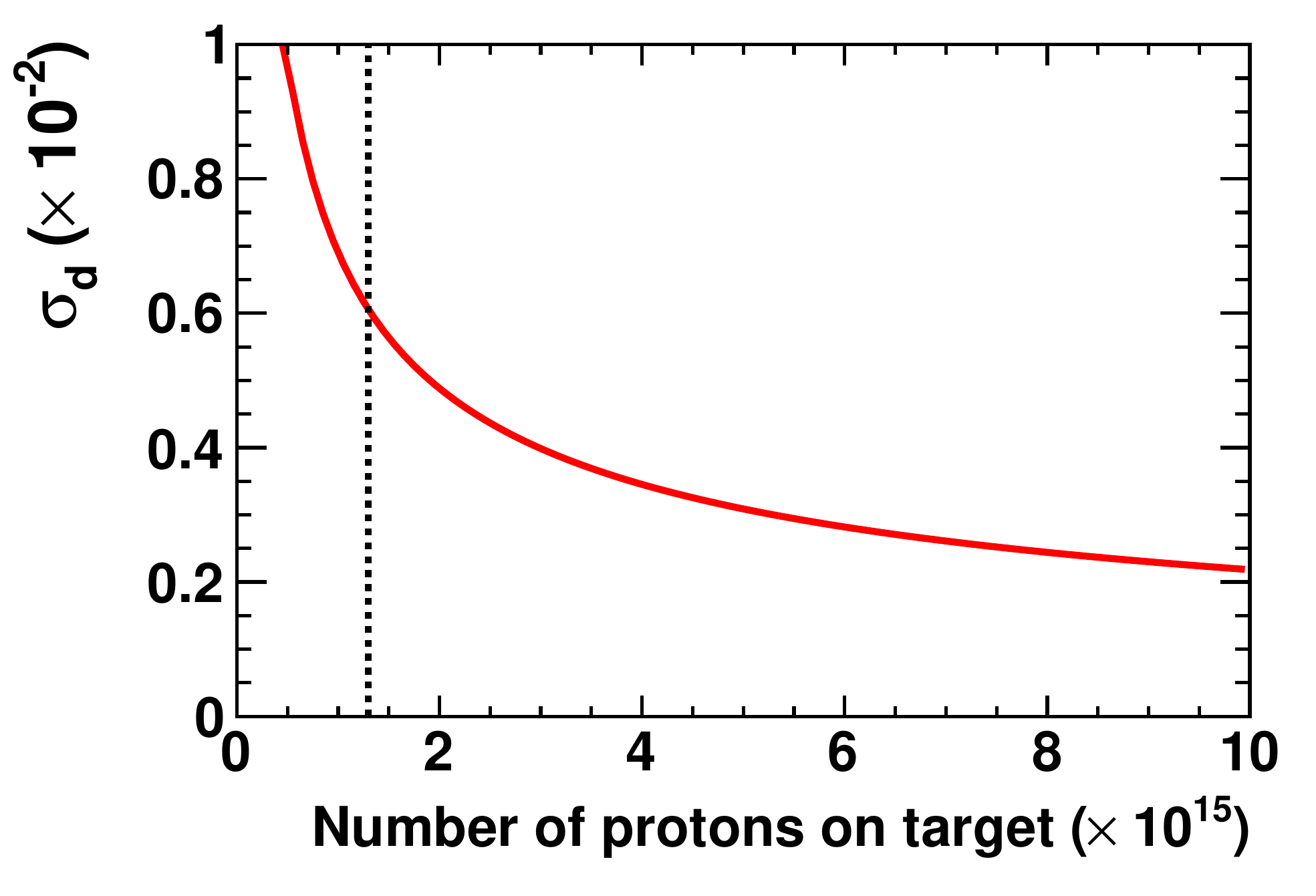}
\includegraphics[width=0.49\textwidth]{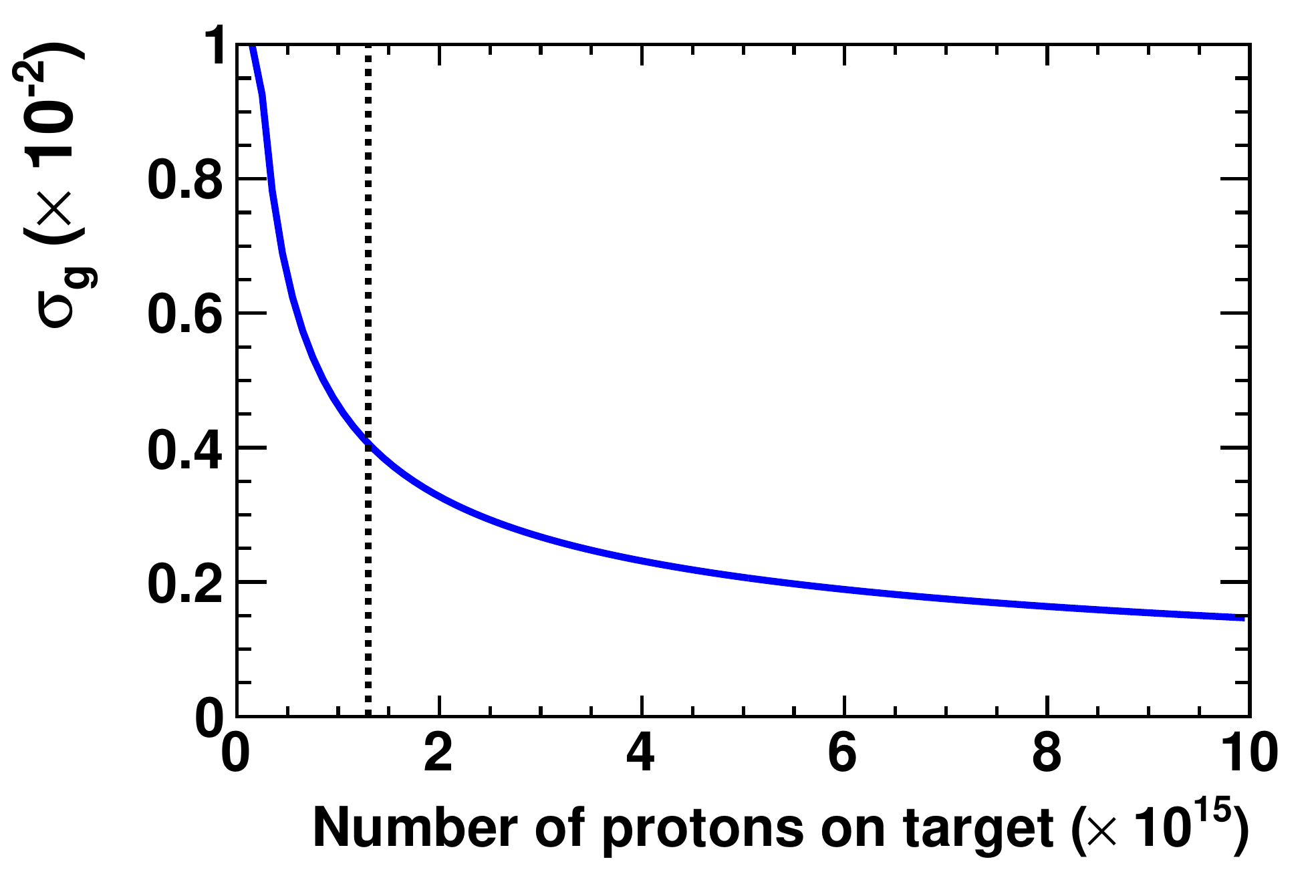}
\caption{Dependence of the (Left) $d$ and (Right) $g$ uncertainties for the
  \Lc baryon, reconstructed in the $\Deltares^{++}\Km$ final state, with the number of protons on target.
  One month of data taking corresponds to $1.3\times 10^{15}$ incident protons (dashed line),
   according to the estimated quantities.
\label{fig:Lambdac_sensitivity}}
\end{figure*}

The dependence of the sensitivity to \Lc EDM and MDM as a function of the number of incident protons on the target is shown in Fig.~\ref{fig:Lambdac_sensitivity}.
The same technique could be applied to any other heavy charged baryon, for instance containing \bquark quark. The production rate is lower than \Lc and the estimates have been studied and discussed in Ref.~\cite{Bagli:2017foe}.

\section{Conclusions}

The unique possibility to search for the EDM of strange and charm baryons at LHC is discussed,
based on the  exploitation of large statistics of baryons with large Lorentz boost and polarization.
The \Lz strange baryons are selected from weak charm baryon decays produced in \proton\proton collisions at
$\approx 14$~\tev center-of-mass energy, while \Lc 
charm baryons are produced in a fixed-target experiment to be installed in the LHC,
in front of the \lhcb detector. Signal events can be reconstructed using the \lhcb detector in
both cases.
The sensitivity to the EDM and  the MDM of the strange and charm baryons arises from the study of the spin precession
in intense electromagnetic fields.
The long-lived \Lz precesses in the magnetic field of the detector tracking system.
Short-lived charm baryons are channeled in a bent crystal attached to the target and the intense electric field
between atomic planes induces the spin precession.   
Sensitivities for the \Lz EDM at the level of $ 1.3 \times 10^{-18}~e\cm$ 
can be achieved
using a data sample corresponding to an integrated luminosity of 50 \invfb to be collected
during the LHC Run 3. A test of \CPT symmetry can be performed by measuring
the MDM of \Lz and \Lbar baryons with a precision of about $4\times 10^{-4}$ on the $g$ factor.
The EDM of the \Lc 
can be searched for with a sensitivity of $2.1\times 10^{-17}\,e\cm$ in 11 days of data taking. 
The proposed experiment would allow about two orders of magnitude improvement in the sensitivity for the
\Lz EDM and the first search  for the charm baryon EDM, 
expanding
the search for new physics
through the EDM of fundamental particles.

\bibliographystyle{ws-procs961x669}
\bibliography{ws-pro-sample,main}

\end{document}